\documentclass[aps,prb,twocolumn,showpacs,preprintnumbers,amsmath,amssymb,superscriptaddress]{revtex4}%
\usepackage{graphicx}% Include figure files
\usepackage{dcolumn}% Align table columns on decimal point
\usepackage{bm}% bold math
\usepackage{color}

\newcommand{\YY}{YBa$_2$Cu$_4$O$_{8}$ }

\begin{document}
%
%[altaffilletter]
%
%\preprint{PREPRINT (\today)}
%
%*******************************************
\title{Finite-size and pressure effects in YBa$_{2}$Cu$_{4}$O$_{8}$
probed by magnetic field penetration depth measurements}
\author{R.~Khasanov}
\email{rustem.khasanov@psi.ch} \altaffiliation[Current address:
]{{\it Laboratory for Neutron Scattering, ETH Z\"urich and Paul
Scherrer Institut, CH-5232 Villigen PSI, Switzerland}}
\altaffiliation{{\it DPMC, Universit\'e de Gen\`eve, 24 Quai
Ernest-Ansermet, 1211 Gen\`eve 4, Switzerland}}
\altaffiliation{{\it Physik-Institut der Universit\"{a}t
Z\"{u}rich, Winterthurerstrasse 190, CH-8057, Switzerland}}
\affiliation{Physik-Institut der Universit\"{a}t Z\"{u}rich,
Winterthurerstrasse 190, CH-8057, Switzerland}
\affiliation{Laboratory for Muon Spin Spectroscopy, Paul Scherrer
Institut, CH-5232 Villigen PSI, Switzerland}
\author{T.~Schneider}
\affiliation{Physik-Institut der Universit\"{a}t Z\"{u}rich,
Winterthurerstrasse 190, CH-8057, Switzerland}
\author{J.~Karpinski}
\affiliation{Solid State Physics Laboratory, ETH 8093 Z\"urich,
Switzerland}
\author{H.~Keller}
\affiliation{Physik-Institut der Universit\"{a}t Z\"{u}rich,
Winterthurerstrasse 190, CH-8057, Switzerland}
%

%now the abstract***************************
\begin{abstract}
We explore the combined pressure and finite-size effects on the
in-plane penetration depth $\lambda _{ab}$ in
YBa$_{2}$Cu$_{4}$O$_{8}$. Even though this cuprate is
stoichiometric the finite-size scaling analysis of $
1/\lambda_{ab}^{2}\left( T\right)$ uncovers the granular nature
and reveals domains with nanoscale size $L_{c}$ along the
$c$-axis. $L_{c}$ ranges from 33.2~\AA\   to 28.9~\AA\   at
pressures from 0.5 to 11.5~kbar. These observations raise serious
doubts on the existence of a phase coherent macroscopic
superconducting state in cuprate superconductors.
\end{abstract}
%*******************************************
%~\\
\maketitle
%\narrowtext
%

%*******************************************
\section{Introduction}
Since the discovery of superconductivity in cuprates by Bednorz
and M\"{u}ller\cite{Bednorz86} a tremendous amount of work has
been devoted to their characterization. Indeed, the issue of
inhomogeneities and their characterization is essential for
applications and the interpretation of experimental data.
Furthermore, there is even increasing evidence that
inhomogeneities are an intrinsic property of cuprates.
\cite{Mesot93,Furrer94,Alekseevskii88,Liu91,Chang92,Cren00,Lang02}
Studies of different cuprate families revealed the segregation of
the material in superconducting and non-superconducting regions.
In particular, neutron scattering experiments provide evidence for
nanoscale cluster formation and percolative superconductivity in
various cuprates.\cite{Mesot93,Furrer94} Electron paramagnetic
resonance (EPR) studies reveal nanoscale phase separation in
superconducting and dielectric regions of
YBa$_2$Cu$_3$O$_{7-\delta}$ with $0.15\leq\delta\leq0.5$.
\cite{Alekseevskii88} Nanoscale spatial variations in the
electronic characteristics have also been observed in underdoped
Bi$_{2}$Sr$_{2}$CaCu$_{2}$O$_{8+\delta }$ with scanning tunnelling
microscopy (STM),\cite{Liu91,Chang92,Cren00,Lang02} while x-ray
diffraction in oxygen doped La$_{2}$CuO$_{4}$ single crystals
\cite {DiCastro00} provide evidence for superconducting domains
with spatial extent $L_{ab}\approx 300$~\AA \  in the $ab$-plane.
Accordingly, there is considerable evidence for granular
superconductivity in the cuprates. Although crystals of the
cuprates are not granular in a structural sense, it occurs when
microscopic superconducting domains are separated by
non-superconducting regions through which they communicate for
instance by Josephson tunnelling to establish the macroscopic
superconducting state.

On the other hand, there is evidence for nearly isolated,
homogeneous and superconducting domains of nanoscale extent
embedded in a non-superconducting matrix.
\cite{Schneider02,Schneider03,Schneider03a,Schneider03b} It stems
from a finite-size scaling analysis of the thermal fluctuation
contributions to the specific heat \cite{Schneider02} and magnetic
penetration depth data.
\cite{Schneider03,Schneider03a,Schneider03b} In an isolated domain
there is a finite-size effect because the correlation length
cannot grow beyond the length of the superconducting domain in
direction $i$. Accordingly there is no sharp phase transition and
the specific heat coefficient will exhibit a blurred peak with a
maximum at $T_{p}<T_{c}$, where $T_{c}$ is the transition
temperature of the homogeneous bulk system. At $T_{p}$ the
correlation length $\xi \left( T\right) $ reaches the limiting
length $L$ of the domain. Similarly, $1/\lambda^{2}$, where
$\lambda$ is the magnetic field penetration depth, does not vanish
at $T_{c}$, but exhibits a tail with an inflection point at
$T_{p}$. This raises serious doubts on the existence of
macroscopic phase coherent superconductivity, suggesting that bulk
superconductivity is achieved by a percolative process. Therefore,
superconducting properties and the spatial extent of the domains
can be probed by thermal fluctuations and the finite-size effects.
This includes the effects of oxygen isotope exchange and pressure
on the domain size. Recently a significant change of spatial
extent of the superconducting domains upon oxygen isotope exchange
has been demonstrated in Y$_{1-x}$Pr$_{x}
$Ba$_{2}$Cu$_{3}$O$_{7-\delta }$.\cite{Schneider03} It revealed
the relevance of local lattice distortions in occurrence of
superconductivity.

This paper addresses the pressure studies of the finite-size
effect in YBa$_2$Cu$_4$O$_8$, which exhibits a rather large and
positive pressure effect (PE) on $T_{c}$, with
$dT_{c}/dp=0.59$~K/kbar.\cite{Scholtz92,Bucher89,VanEinige90} Even
though this cuprate is {\it stoichiometric} the finite-size
scaling analysis uncovers the existence of {\it nanoscale domains}
with a spatial extent $L_{c}$ along the crystallographic $c$-axis.
The value of $L_c$ decreases from $33.2$ to $28.9$~\AA \  with
increasing pressure from $0.5$ to $11.5$~kbar. Accordingly,
$T_{c}$ {\it increases} with {\it reduced thickness} $L_{c}$ of
the domains.

The paper is organized as follows. In
Sec.~\ref{seq:Theoretical_background} we sketch the finite-size
scaling theory adapted for the analysis of penetration depth data.
In Sec.~\ref{seq:Experimental} we describe sample preparation
procedure and the experimental technique, adopted to deduce the
in-plane magnetic field penetration depth $\lambda _{ab}$ from the
Meissner fraction measurements. In
Sec.~\ref{seq:The_finite-size_analysis} we perform the finite-size
analysis of the data for the in-plane magnetic penetration depth
$\lambda _{ab}\left( T\right) $ taken at different pressures.
Sec.~\ref{seq:pressure_dependence_domain_size} comprises the
analysis of the pressure dependence of the domain lengths $L_{c}$
along the crystallographic $c$-axis.

\section{Theoretical background} \label{seq:Theoretical_background}

In a homogeneous bulk system, undergoing a fluctuation dominated
continuous phase transition at $T_{c}$, the correlation length
diverges as \cite{Shan-Keng76}
\begin{equation}
\xi (T)=\xi _{0}^{\pm }|T/T_{c}-1|^{-\nu }=\xi _{0}^{\pm
}|t|^{-\nu }
 \label{eq:critical_exponent}
\end{equation}
($\pm $ refers to $T>T_{c}$ and $T<T_{c}$, respectively, $\xi
_{0}^{\pm }$ is the critical amplitude and $\nu $ is the
associated critical exponent). There is mounting evidence that in
the experimentally accessible critical regime cuprates belong to
the 3D-XY universality class (like superfluid He$^{4}$) with $\nu
\approx 2/3$.\cite{Schneider00} Since the order parameter $\Psi$
is a complex scalar it corresponds to a vector with two
components. For this reason below $T_{c}$ there are  two
correlation lengths. The longitudinal one, $\xi ^{l}$, is
associated with the correlation function $\left\langle Re\Psi
\left( R\right) Re\Psi \left( 0\right) \right\rangle -\left\langle
Re\Psi \left( 0\right) \right\rangle ^{2}$, while the transverse
one, $\xi ^{t}$, measures the decay of $\left\langle Im\Psi \left(
R\right) Im\Psi \left( 0\right) \right\rangle $. In the long
wavelength limit considered here, the total correlation function
is dominated by transverse fluctuations so that $\xi ^{t}$ is the
relevant length scale.\cite{Schneider00,Hohenberg76} Suppose that
the cuprates are granular, consisting of superconducting domains
embedded in a non-superconducting matrix. Denoting the spatial
extent of the domains along the crystallographic $a$, $b$ and $c$
-axis with $L_{a}$, $L_{b}$ and $L_{c}$, the transverse
correlation lengths $ \xi _{i}^{t}$ cannot diverge according to
Eq.~\ref{eq:critical_exponent} but are limited by
\begin{equation}
\xi _{i}^{t}\xi _{j}^{t}\leq L_{k}^{2},\ i\neq j\neq k.
\label{eq:correlation_limit}
\end{equation}
Consequently, for finite superconducting domains, the
thermodynamic quantities, like the specific heat and penetration
depth, are smooth functions of temperature. As a remnant of the
singularity at $T_{c}$ these quantities exhibit a so called
finite-size effect,\cite{Fisher72,Cardy88} namely a maximum or an
inflection point at $T_{p_{i}}$
\begin{equation}
\xi _{i}^{t}(T_{p_{i}})\xi _{j}^{t}(T_{p_{i}})=L_{k}^{2},\ i\neq
j\neq k. \label{eq:correlation_limit_inf_point}
\end{equation}

Close to criticality of the infinite system, the thermodynamic
properties of its finite counterpart are well described by the
finite-size scaling theory.\cite{Fisher72,Cardy88} In particular,
the thermodynamic observable $Q$ adopts the scaling
form\cite{Schultka95}
%
%\textcolor{red}{ If the system is confined in a finite geometry
%the finite-size scaling theory \cite{Fisher72} is usually
%describing rather well the behavior of the system at temperatures
%close to the transition one.  In this case the physical quantity
%$Q$ adopts the scaling form\cite{Schultka95}}
%
\begin{equation}
\frac{Q\left( t,L\right) }{Q\left( t,L=\infty \right) }=f\left(
x\right) ,\ x=\frac{L}{\xi \left( t,L=\infty \right) }.  \label{O}
\end{equation}
The scaling function $f$ depends only on the dimensionless ratio
$L/\xi \left( t,L=\infty \right) $ and does not depend on
microscopic details of the system. It does, however, depend on the
boundary conditions and the geometry of the system. As an example,
for $Q(t,L)=\xi _{i}^{t}\xi_{j}^{t}$ we obtain from
Eqs.~(\ref{eq:correlation_limit_inf_point}) and (\ref{O}) the
finite-size scaling relation
\begin{equation}
\frac{\xi _{i}^{t}\xi _{j}^{t}}{\xi _{0i}^{t}\xi
_{0j}^{t}}|t|^{2\nu
}=f\left( \frac{sign(t)|t|^{\nu }L_{k}}{\sqrt{\xi _{0i}^{t}\xi _{0j}^{t}}}%
\right) .
 \label{scaling-finction_coherence_length}
\end{equation}
To relate this combination of transverse correlation lengths to an
experimentally accessible quantity we invoke the universal
relation \cite{Schneider00,Schneider98}
\begin{equation}
\frac{1}{\lambda _{i}^{2}(T)}=\frac{16\pi ^{3}k_{B}T}{\Phi
_{0}^{2}\xi _{i}^{t}(T)},
 \label{eq:universal relation}
\end{equation}
which holds in the 3D-XY universality class ($\Phi _{0}$ is the
flux quantum, $k_{B}$ is the Boltzmann's constant and
$\lambda_{i}$ is the London penetration depth). In this case the
Eq.~(\ref {scaling-finction_coherence_length}) reduces to
\begin{equation}
\frac{\lambda _{0i}\lambda _{0j}}{\lambda _{i}\lambda
_{j}}|t|^{-\nu }=\left( \frac{\xi _{i}^{t}\xi _{j}^{t}}{\xi
_{0i}^{t}\xi _{0j}^{t}}\right) ^{-\frac{1}{2}}|t|^{-\nu }=g(y),
\label{eq:scaling-finction_lambda_1}
\end{equation}
where $y$ is equal to
\begin{equation}
y=sign\left( t\right) \left| t\right| \left(
\frac{L_{k}}{\sqrt{\xi
_{0i}^{t}\xi _{0j}^{t}}}\right) ^{1/\nu }=sign\left( t\right) \left| \frac{t%
}{t_{p_{k}}}\right|
 \nonumber
\end{equation}

For the homogenous system ($L_{k}\rightarrow \infty $) and $t\neq
0$ ($y\neq 0$), $g(y)$ corresponds to the stepwise function
\begin{equation}
g_{\infty }\left( y<0\right) =1,\ g_{\infty }\left( y>0\right) =0.
 \label{eq10c}
\end{equation}
While for the system confined by the finite geometry ($L_{k}\neq
0$) the scaling function $g(y)$ diverges at $t\rightarrow 0$
($y\rightarrow 0$) as
\begin{equation}
g\left( y\rightarrow 0\right) =g_{0k}y^{-\nu }=g_{0k}\left( \left| \frac{t}{%
t_{p_{k}}}\right| \right) ^{-\nu }.
 \label{eq:10d}
\end{equation}

\bigskip

\bigskip

In order to obtain the absolute values of the inflection
temperature $T_{p_i}$ and the superconducting domain size $L_i$
one can use the following procedure. Combining
Eqs.~(\ref{eq:correlation_limit_inf_point}) and (\ref{eq:universal
relation}) one obtains at the inflection point
\begin{equation}
\left. \frac{1}{\lambda _{i}(T)\lambda _{j}(T)}\right| _{T=T_{p_{k}}}=\frac{%
16\pi ^{3}k_{B}T_{p_{k}}}{\Phi _{0}^{2}}\frac{1}{L_{k}}.
\label{eq4}
\end{equation}
For an infinite and homogeneous system $1/\left( \lambda
_{i}\left( T\right) \lambda _{j}\left( T\right) \right) $
decreases continuously with increasing temperature and vanishes at
$T_{c}$, while for finite domains it does not vanish and exhibits
an inflection point at $T_{p_{k}}<T_{c}$, so that
\begin{equation}
\left. d\left( \frac{1}{\lambda _{i}\left( T\right) \lambda
_{j}\left( T\right) }\right) /dT\right| _{T=T_{p_{k}}}=extremum
 \label{eq5}
\end{equation}
Note, that in this paper we analyze experimental data for the
temperature dependence of the in-plane penetration depth
$\lambda_{ab}$ taken at various applied pressures. In this case
the domain size along $c$-axis ($L_{c}$) can be estimated
according to Eq.~(\ref{eq4}) as
\begin{equation}
L_{c}=\frac{16\pi ^{3}k_{B}T_{p_{c}}\left( \lambda _{a}\left(
T\right) \lambda _{b}\left( T\right) \right) _{T=T_{p_{c}}}}{\Phi
_{0}^{2}},
 \label{eq:Lc_one}
\end{equation}
that for $\lambda_a\simeq\lambda_b$ reduces to
%so that for $\lambda _{a}\simeq \lambda _{b}$
%
\begin{equation}
L_{c}\simeq \frac{16\pi ^{3}k_{B}T_{p_{c}}\lambda _{ab}^{2}\left(
T_{p_{c}}\right) }{\Phi _{0}^{2}}.
 \label{eq:Lc_two}
\end{equation}

To summarize, the main signatures for the existence of finite-size
behavior appearing in the temperature dependence of $\lambda
_{ab}^{2}$ are:

\begin{itemize}
\item[(i)]  The scaling function $g(y)$ diverges at $T=T_{c}$
$(t=0,$ $y=0)$ [Eq.~(\ref{eq:10d})].

\item[(ii)]  $\lambda ^{-2}(T)$ has an inflection point at
$T_{p}<T_{c}$ [Eq.(\ref{eq4})].

\item[(iii)]  The first derivative of $\lambda ^{-2}(T)$ at
$T=T_{p}$ has an extremum [Eq.(\ref{eq5})].
\end{itemize}

\section{Experimental Details}\label{seq:Experimental}
\subsection{Sample preparation and characterization}

The polycrystalline YBa$_{2}$Cu$_{4}$O$_{8}$ samples were
synthesized by solid-state reactions using high-purity
Y$_{2}$O$_{3}$, BaCO$_{3}$ and CuO. The samples were calcinated at
$880-935$~$^{o}$C in air for $110$~hours with several intermediate
grindings. The phase-purity of the material was examined using a
powder x-ray diffractometer. Only YBa$_{2}$Cu$_{3}$O$_{7-x}$ and
CuO phases were revealed. The synthesis was continued at high
oxygen pressure of $500$~bar, at $1000$~$^{o}$C during 30~hours.
The x-ray diffraction measurements performed after the final stage
of the synthesis revealed 95~\% of YBa$_{2}$Cu$_{4}$O$_{8}$ phase.
The sample was then regrounded in a mortar for about $60$~min in
order to obtain sufficiently small grains, as required for the
determination of $\lambda$ from Meissner fraction measurements.
The field-cooled (FC) magnetization ($M$) measurements were
performed with a Quantum Design SQUID magnetometer in a field of $
0.5$~mT for temperatures ranging from $5$~K to $100$~K. The
absence of weak links between grains has been confirmed by the
linear magnetic field dependence of the FC magnetization, measured
at $0.5$~mT, $1$~mT and $1.5$~mT for each pressure at $T=10$~K.

The hydrostatic pressure was generated in a copper-beryllium
piston cylinder clamp that was especially designed for
magnetization under pressure measurements (see
Ref.~[\onlinecite{Straessle02}]). The sample was mounted in a led
container filled with Fluorient FC77 as a pressure transmitting
medium with a sample to liquid volume ratio approximately $1/8$.
The pressure was measured in situ at 7~K by using the $T_{c}$
shift of the led container.

\subsection{Determination of the temperature dependence of
$\lambda$ from Meissner fraction measurements}

\begin{figure}[htb]
%\centering
\includegraphics[width=1.0\linewidth]{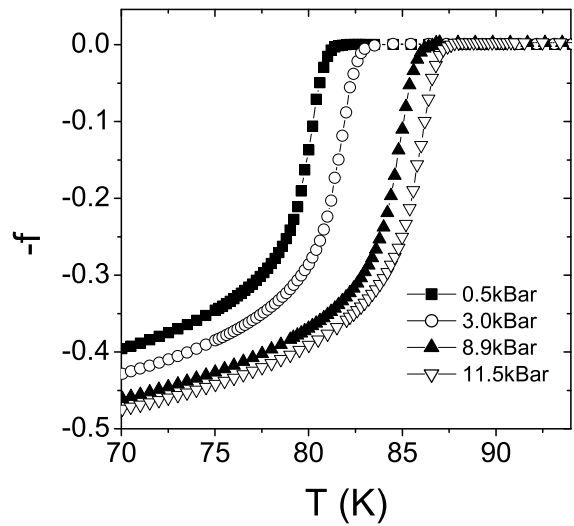}
 \vspace{-1.0cm}
\caption{The temperature dependence of the Meissner fraction $f$
obtained from low-filed (0.5~mT, FC ) magnetization measurements
for various pressures.}
\label{Magnetization}
\end{figure}

The temperature dependence of $\lambda ^{-2}$ was extracted from
the Meissner fraction $f$ \ deduced from low-field (0.5~mT FC)
magnetization data using the relation:\cite{Blundel01}
\begin{equation}
f(T)=\left( \frac{H}{M(T)}-N\right) ^{-1},
\end{equation}
where $H$ denotes the external magnetic field and $N$ is the
demagnetization factor. $N=1/3$ was taken assuming that the sample
grains are spherical. Fig.~\ref{Magnetization} shows the
temperature dependence of the Meissner fraction close to $T_{c}$
for different pressures. Three important features emerge: (i) The
transition temperature $ T_{c}$ increases with increasing
pressure. The value of $dT_c/dp= 0.59$~K/kbar is found, which is
in good agreement with the literature data.
\cite{Scholtz92,Bucher89,VanEinige90} (ii) The value of $f$
(Fig.~\ref{Magnetization}) is much smaller than 1, confirming that
the average grain size of the sample is compatible with $\lambda$.
The reduction of $f$ is caused by the field penetration at the
surface of each individual grain for distances of the order of
$\lambda $.\cite{Zhao97} (iii) The absolute value of the Meissner
fraction increases with pressure. Since the average grain size
does not change under pressure and the grains are decoupled from
each other, the rise of $f$ must be attributed to a decrease of
the magnetic penetration depth $\lambda $.

\begin{figure}[htb]
%\centering
\includegraphics[width=1.0\linewidth]{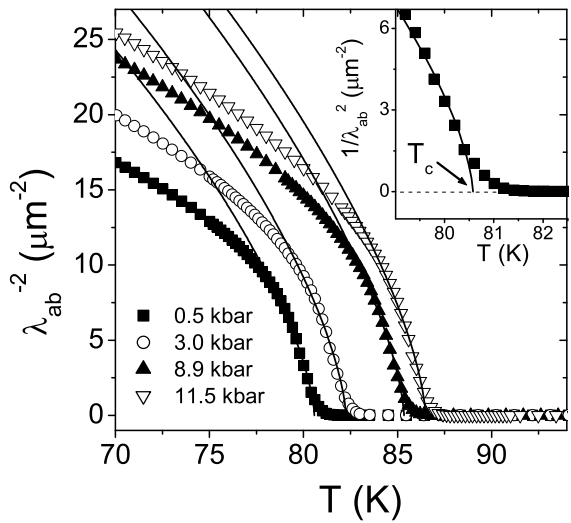}
 \vspace{-1.0cm}
\caption{The temperature dependence of $\lambda_{ab}^{-2}$ for
various pressures obtained from $f(T)$ data (see
Fig.~\ref{Magnetization}) by using Eq.~(\ref{eq:Shoenberg}). The
solid lines indicate the leading critical behavior of a homogenous
bulk system according to
Eq.~(\ref{eq:lambda_leading_critical_behavior}) with the
parameters listed in Table~\ref{Table1}. The inset shows
$\lambda_{ab}^{-2}(T)$ in the vicinity of $T_c$ for $p=0.5$~kbar.
The deviation of data points from theoretical curves clearly
indicates the finite-size behavior of the system. }
\label{fig:lambda}
\end{figure}

The temperature dependence of $\lambda$ was analyzed on the basis
of model suggested by Shoenberg.\cite{Shoenberg40} According to
Ref.~[\onlinecite{Shoenberg40}]  the temperature dependence of the
Meissner fraction is given by
\begin{equation}
f(T)=1-3\left( \frac{\lambda (T)}{R}\right) \coth {\left(
\frac{R}{\lambda (T)}\right) }+3\left( \frac{\lambda
(T)}{R}\right) ^{2},
 \label{eq:Shoenberg}
\end{equation}
where $2R$ is the average grain diameter. By solving this
nonlinear equation, $\lambda$ for each value of $f$ was extracted,
and with that the whole temperature dependencies of $\lambda$ was
reconstructed. Since the sample consists of an anisotropic
non-oriented powder, the extracted $\lambda$ is the so called
effective penetration depth $\lambda _{eff}$ (powder average).
However, for sufficiently anisotropic extreme type II
superconductors, including YBa$_{2}$Cu$_{4}$O$_{8}$, $\lambda
_{eff}$ is proportional to the in-plane penetration depth\ in
terms of $\lambda _{eff}=1.31\lambda _{ab}$.\cite{Fesenko91} The
resulting temperature dependencies of $\lambda _{ab}^{-2}$
evaluated at different pressures are depicted in
Fig.~\ref{fig:lambda}. Due to the unknown average grain size the
data in Fig.~\ref{fig:lambda} are normalized to the
$\lambda_{ab}^{-2}(0)$, taken from $\mu $SR measurements.
\cite{Khasanov04} The solid lines indicate the leading critical
behavior of $\lambda _{ab}^{-2}(T)$ for homogenous and infinite
domains
\begin{equation}
\lambda _{ab}^{-2}(T)=\lambda _{0ab}^{-2}|t|^{\nu },\ \ \nu =2/3
\label{eq:lambda_leading_critical_behavior}
\end{equation}
This equation is a consequence of
Eqs.~(\ref{eq:critical_exponent}) and (\ref {eq:universal
relation}) with critical amplitude
\begin{equation}
\lambda _{0ab}^{-2}=\frac{16\pi ^{3}k_{B}T_{c}}{\Phi _{0}^{2}\xi
_{0ab}^{t}}.
\end{equation}
The values of $\lambda _{0ab}$ and $T_{c}$ obtained from the fit
of Eq.~(\ref{eq:lambda_leading_critical_behavior}) to the
experimental data presented in Fig.~\ref{fig:lambda} are
summarized in Table ~\ref{Table1}. Since
Eq.~(\ref{eq:lambda_leading_critical_behavior}) is valid in the
vicinity of $T_{c}$ only, we restricted the fit to the interval
from $T_{c}$ to $T_{c}-3$~K.

\section{The finite-size analysis} \label{seq:The_finite-size_analysis}

\begin{figure}[htb]
%\centering
\includegraphics[width=1.0\linewidth]{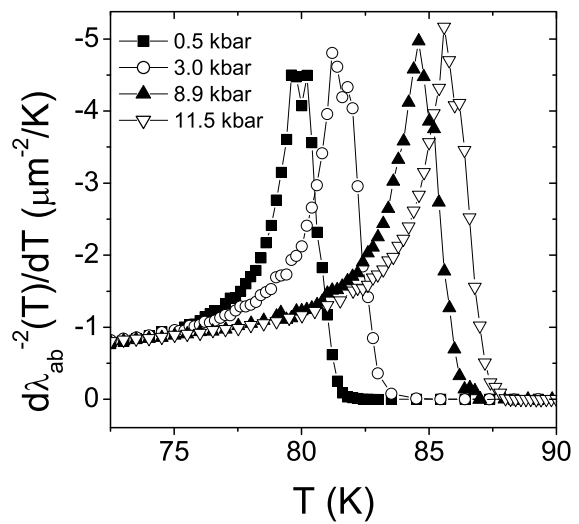}
 \vspace{-1.0cm}
\caption{$d\lambda_{ab}^{-2}(T)/dT$ vs. $T$ at different
pressures. The maximum of $d\lambda_{ab}^{-2}(T)/dT$ is at a
$T=T_{p_c}$.}
\label{fig:derivative_lambda}
\end{figure}

The essential characteristic of a homogeneous bulk cuprate
superconductor is a sharp superconductor to normal state
transition. A glance to the inset of Fig.~\ref{fig:lambda} shows
that in the samples considered here that is not the case. The
transition occurs smoothly and there is a tail pointing to a
finite-size effect associated with an inflection point at some
characteristic temperature $T_{p}$. As outlined above [Eq.~(\ref
{eq:correlation_limit_inf_point})] at $T_{p}$ the transverse
correlation length $\xi _{ab}^{t}$ attains the limiting length
along the $c$-axis. To substantiate the occurrence of an
inflection point we show in Fig.~\ref{fig:derivative_lambda}
$d\lambda _{ab}^{-2}(T)/dT$ versus $T$ for
YBa$_{2}$Cu$_{4}$O$_{8}$ samples at different pressures. The
extreme in the first derivative of $\lambda _{ab}^{-2}(T)$ clearly
reveals the existence of an inflection point at $T_{p_{c}}<T_{c}$.
The absolute values of $T_{p_c}$ obtained from a parabolic fit to
experimental data around maximum point are summarized in
Table~\ref{Table1}.

\begin{figure}[htb]
%\centering
\includegraphics[width=1.0\linewidth]{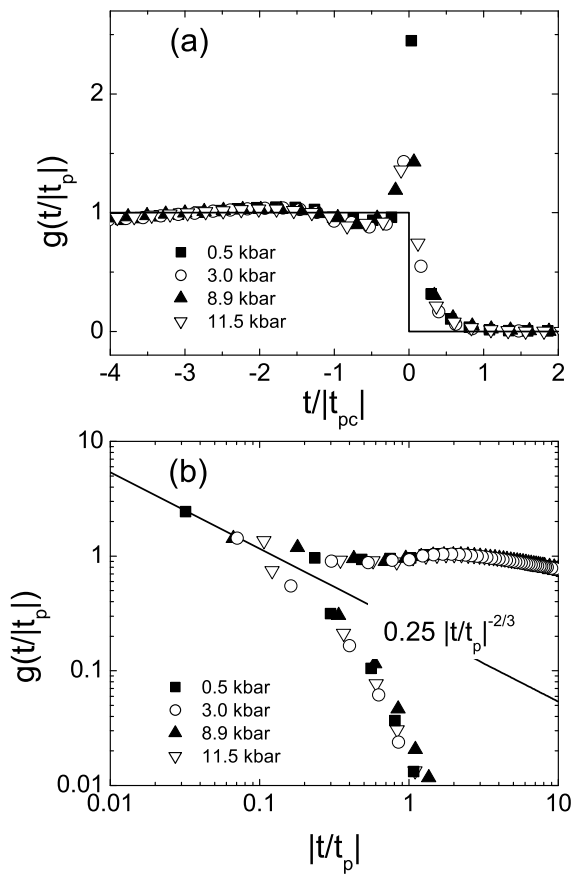}
 \vspace{-1.0cm}
\caption{Finite size scaling function $g$
[Eq.~(\ref{eq:scaling-finction_lambda})] versus $t/\left|
t_{p_{c}}\right| $ (a) and versus $\left|t/ t_{p_{c}}\right|$ in a
logarithmic scale (b) for different pressures. The solid stepwise
line in (a) is the $g_\infty$ function for the homogenous
($L_c\rightarrow\infty$) system [see Eq.~(\ref{eq10c})]. The solid
line in (b) is the leading finite-size behavior
$g_{0c}|t/t_{p_c}|^{-2/3}$ at $t\rightarrow 0$ with $g_{0c}=0.25$
as it follows from Eq.~(\ref{eq:10d}). }
\label{fig:scaling_function}
\end{figure}

To substantiate the finite-size scenario further, we explore the
scaling properties of the data with respect to the consistency
with the finite-size scaling function. Noting that, $\lambda
_{a}\simeq \lambda _{b}$ and $\nu \approx 2/3$ in the 3D-XY
universality class, Eq.(\ref {eq:scaling-finction_lambda_1}) can
be rewritten as
\begin{equation}
\frac{\lambda _{0ab}^{2}}{\lambda _{ab}^{2}}|t|^{-2/3}=g\left( \frac{t}{%
|t_{p_{c}}|}\right) .
 \label{eq:scaling-finction_lambda}
\end{equation}
Figure~\ref{fig:scaling_function}(a) shows the resulting scaling
function. For comparison, we included the limiting behavior of the
finite-size scaling function $g_\infty$ [see Eq.~(\ref {eq10c})]
for a homogeneous ($L_c\rightarrow \infty $) system [the solid
line in Fig.~\ref{fig:scaling_function}(a)]. As it is seen, there
is quite a good agreement between the experimental data and
$g_\infty$ function for $t/t_{p_c}$ far from 0 ($T$ far from
$T_c$), whereas in the vicinity of 0 ($T\sim T_c$) the data are
completely inconsistent with such a stepwise behavior. The
experimental data have to diverge at $t/t_{p_c}\rightarrow 0$. As
outlined in Sec.~\ref{seq:Theoretical_background} divergence of
the scaling function $g(t/|t_{p_c}|)$ at $t/t_{p_c}\rightarrow 0$
implies that the system is confined by a finite geometry. To
strengthen this point we display in
Fig.~\ref{fig:scaling_function}(b) the comparison with the leading
finite-size behavior $g_{c}(t\rightarrow
0)=g_{0c}|t/t_{p_{c}}|^{-2/3}$, as it is follows from
Eq.~(\ref{eq:10d}). Noting that the amplitude $g_{0c}$ depends on
the shape of the domains and on the boundary conditions,
\cite{Schneider03b} it is remarkable that the experimental data
collapses on two branches. Accordingly, the shape of the domains
and the boundary conditions on the interface do not change
significantly under applied pressure. The straight line in
Fig.~\ref {fig:scaling_function} (b) corresponds to $g_{0c}$
$\simeq 0.25$. This value is compatible with $g_{0_{c}}\simeq 0.5$
found for YBa$_{2}$Cu$_{3}$O$_{6.7}$ oriented powder and for
Bi$_{2}$Sr$_{2}$CaCu$_{2}$O$_{8+\delta }$ thin films.
\cite{Schneider03b} Much larger values, $g_{0c}\approx 1.1-1.6$,
have been found for Bi$_{2}$Sr$_{2}$CaCu$_{2}$O$_{8+\delta }$
single crystals \cite{Schneider03b} and for
Y$_{1-x}$Pr$_{x}$Ba$_{2}$Cu$_{3}$O$_{7-\delta }$ powders with
$0.0\leq x\leq0.3$.\cite{Schneider03}

\begin{table*}[htb]
 \caption[~]{Finite size estimates for $T_{p_{c}}$, $\lambda_{0ab}^{-2}$,
$\lambda_{ab}^{-2}$($T_{p_c}$) and the resulting relative shifts
$\Delta T_{p_{c}}/T_{p_{c}}$and $\Delta \lambda _{ab}^{2} \left(
T_{p_{c}}\right) / \lambda _{ab}^{2} \left( T_{p_{c}} \right) $
for different pressures. $L_{p_{c}}$ and $\Delta
L_{p_{c}}/L_{p_{c}} $ are deduced from Eq.(\ref{eq:Lc_two}). }
 \label{Table1}
\begin{tabular}{c|ccccccccccc}
 \hline
 \hline
 &$T_c$&$T_{p_c}$&$\lambda_{0ab}^{-2}$&$\lambda_{ab}^{-2}$($T_{p_c}$)&$L_c$&
 $\frac{\Delta T_c}{T_c}$&$\frac{\Delta T_{p_c}}{T_{p_c}}$&
 $\frac{\Delta\lambda_{ab}^{-2}(T_{p_c})}{\lambda_{ab}^{-2}(T_{p_c})}$&
 $\frac{\Delta L_c}{L_c}$\\
 &(K)&(K)&($\mu$m$^{-2}$)&($\mu$m$^{-2}$)&(\AA)&(\%)&(\%)&(\%)&(\%)\\
 \hline
 0.5~kbar&80.58(4)&79.85(3)&93.8(1.5)&3.85(12)&33.2(9)&-&-&-&-\\
 3.0~kbar&82.26(3)&81.45(3)&100.6(1.5)&4.29(11)&30.5(6)&2.08(6)&2.00(5)&11.3(3.7)&-8.2(3.7)\\
 8.9~kbar&85.34(3)&84.55(3)&108.0(1.4)&4.60(13)&29.3(8)&5.91(6)&5.89(5)&19.7(4.0)&-11.7(4.0)\\
 11.5~kbar&86.49(3)&85.65(3)&109.8(1.5)&4.74(13)&28.9(6)&7.33(6)&7.26(5)&23.3(3.8)&-12.8(3.7)\\

% p~(kbar) & 0.5 & 3.0 & 8.9 & 11.5 \\
% \hline
% $T_c$ & 80.58(4) & 82.26(3) & 85.34(3) & 86.49(3) \\
% $T_{p_c}$~(K)& 79.85(3) & 81.45(3)  & 84.55(3) & 85.65(3) \\
% $t_{p_c}$ &-0.0091(6)&-0.0098(5)&-0.0093(5)&-0.0097(5)\\
% $\lambda_{0ab}^{-2}$~($\mu$m$^{-2}$) & 121.6(1.9) & 130.1(2.0) & 139.6(1.8) & 141.9(2.0) \\
% $\lambda_{ab}^{-2}$($T_{p_c}$)~($\mu$m$^{-2}$)& 4.97(15) & 5.53(14) & 5.95(17) & 6.13(16) \\
% $L_c$~(\AA)&25.7(7)&23.6(5)&22.7(6)&22.4(5)\\
% $\Delta T_c/T_c$~(\%)&-&2.08(6)&5.91(6)&7.33(6)\\
% $\Delta T_{p_c}/T_{p_c}$~(\%)&-&2.00(5)&5.89(5)&7.26(5)\\
% $\Delta\lambda_{ab}^{-2}(T_{p_c})/\lambda_{ab}^{-2}(T_{p_c})$~(\%)&-&11.3(3.7)&19.7(4.0)&23.3(3.8)\\
% $\Delta L_c/L_c$~(\%)&-&-8.2(3.7)&-11.7(4.0)&-12.8(3.7)\\
 \hline
 \hline
\end{tabular}
\end{table*}

To summarize, the finite-size scaling analysis of the in-plane
penetration depth data for YBa$_{2}$Cu$_{4}$O$_{8}$ is fully
consistent with a finite-size effect. Indeed we established the
consistency with all three characteristics of a finite-size effect
(see Sec.~\ref {seq:Theoretical_background}). The finite-size
estimates for  $\lambda_{0ab}^{-2}$ and
$\lambda_{ab}^{-2}$($T_{p_c}$) at different pressures  are
summarized in Table~\ref{Table1}.

\section{The pressure dependence of the domain size } \label{seq:pressure_dependence_domain_size}

\begin{figure}[htb]
%\centering
\includegraphics[width=1.0\linewidth]{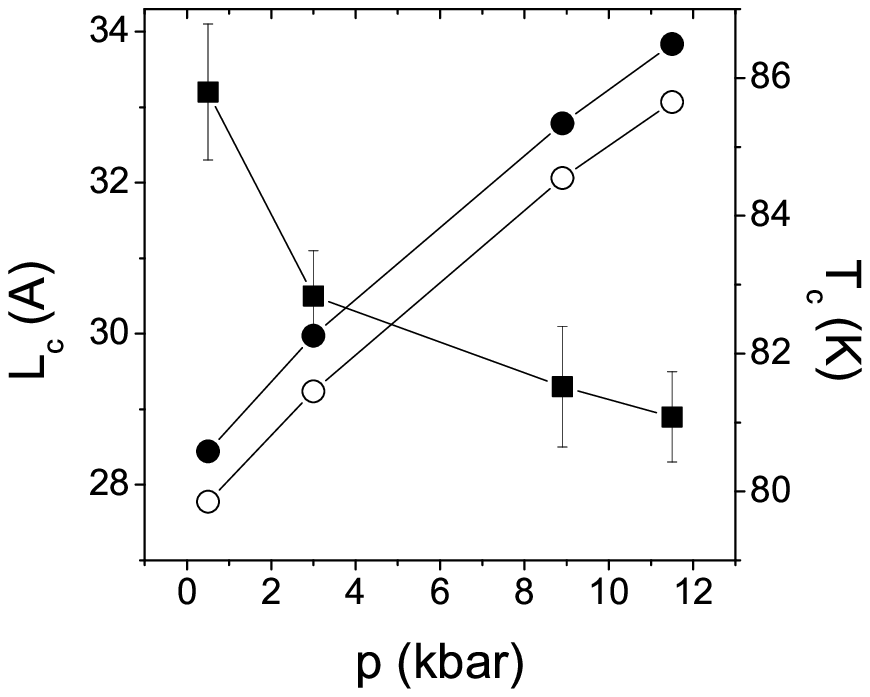}
 \vspace{-1.0cm}
\caption{The pressure dependence of the domain size along the
$c$-axis $L_c$ ($\blacksquare$), transition temperature $T_c$
($\bullet$), and inflection temperature $T_{p_c}$ ($\circ$) in
\YY. }
\label{fig:Lc_vs_pressure}
\end{figure}

Using Eq.~(\ref{eq:Lc_two}) and the estimates for $T_{p_{c}}$ and
$\lambda _{ab}^{-2}(T_{p_{c}})$ listed in Table~\ref{Table1}, the
domain size $L_{c} $ along the $c$-axis is readily calculated. In
Fig.~\ref{fig:Lc_vs_pressure} we display the pressure dependence
of $L_{c}$, $T_{c}(p)$ and $T_{p_{c}}(p) $.
% for the data points listed in Table~\ref{Table1}.
It is seen that $L_{c}$ decreases with pressure, whereas
$T_{c}(p)$ and $T_{p_{c}}(p)$ increase almost linearly. On the
other hand $T_{c}$ and $T_{p_{c}}$ increase with decreasing
$L_{c}$. This agrees with the behavior found in
Y$_{1-x}$Pr$_{x}$Ba$_{2}$Cu$_{3}$O$_{7-\delta }$, where $T_{c}$
and $ T_{p_{c}}$ were found to increase with reduced $L_{c}$.
\cite{Schneider03} Here $T_{c}$ and $T_{p_{c}}$ have been reduced
by increasing the Pr content $x$. In this context it is
interesting to note that in granular aluminium  $T_{c}$ was found
to increase with reduced grain size.\cite{deutscher2} Another
striking feature is the nanoscale magnitude of $L_{c}$, even
though YBa$_{2}$Cu$_{4}$O$_{8}$ is stoichiometric.

\begin{figure}[htb]
%\centering
\includegraphics[width=1.0\linewidth]{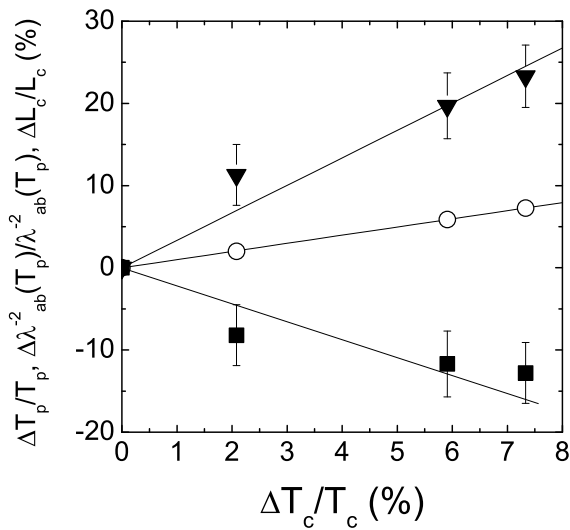}
 \vspace{-1.0cm}
\caption{The relative shifts of  $\Delta \protect\lambda
_{ab}^{-2}(T_{p_{c}})/\protect \lambda _{ab}^{-2}(T_{p_{c}})$
($\blacktriangledown$);  $\Delta L_{c}/L_{c}$ ($ \blacksquare $),
and $\Delta T_{p_{c}}/T_{p_{c}}$ ($ \bigcirc  $) vs the relative
shift of $\Delta T_{c}/T_c$. The solid lines represent the linear
fits with $\Delta L_{c}/L_{c}= -2.18(26)\cdot\Delta T_{c}/T_{c}$,
$\Delta T_{p_{c}}/T_{p_{c}}=0.99(1)\cdot \Delta T_{c}/T_{c}$ and $
\Delta \lambda _{ab}^{-2}\left( T_{p_{c}}\right) /\lambda
_{ab}^{-2}\left( T_{p_{c}}\right) = 3.34(27)\cdot\Delta
T_{c}/T_{c}$.  }
\label{fig:finite-size_shifts}
\end{figure}

To check the consistency of our estimates we plot in Fig.~\ref
{fig:finite-size_shifts} the relative shifts of $T_{p_{c}}$,
$\lambda _{ab}^{2}\left( T_{p_{c}}\right) $ and $L_{c}$ versus the
relative shift of $T_{c}$. According to Eq.(\ref{eq4}) these
shifts are not independent but related by
\begin{equation}
\frac{\Delta L_{c}}{L_{c}}=\frac{\Delta
T_{p_{c}}}{T_{p_{c}}}-\frac{\Delta \lambda _{ab}^{-2}\left(
T_{p_{c}}\right) }{\lambda _{ab}^{-2}\left( T_{p_{c}}\right) }.
\label{eq:relative_shifts}
\end{equation}
The straight lines correspond to the linear fits with $\Delta
L_{c}/L_{c}= -2.18(26)\cdot\Delta T_{c}/T_{c}$, $\Delta
T_{p_{c}}/T_{p_{c}}=0.99(1)\cdot \Delta T_{c}/T_{c}$ and $ \Delta
\lambda _{ab}^{-2}\left( T_{p_{c}}\right) /\lambda
_{ab}^{-2}\left( T_{p_{c}}\right) = 3.34(27)\cdot\Delta
T_{c}/T_{c}$, revealing that Eq.~(\ref{eq:relative_shifts}) is
well satisfied. Furthermore, these estimates show that the
reduction of $L_{c}$ with pressure reflects the fact that the
pressure effect on $\lambda _{ab}^{-2}(T_{p_{c}})$ exceeds the
effect on $ T_{c}$ considerably. Having established the
consistency of our estimates it is essential to recognize that the
occurrence of nanoscale superconducting domains is not an artefact
of YBa$_{2}$Cu$_{4}$O$_{8}$ and our samples. Indeed the existence
of nanoscale domains has been established in a variety of
cuprates, including films, powders and single
crystals.\cite{Schneider03b}

\section{Conclusion}

To summarize, we report the first observation of  combined
finite-size and pressure effects on the lengths $L_{c}$ of the
superconducting domains along the $c$-axis and the  in-plane
penetration depth $\lambda _{ab}$ in a cuprate superconductor. The
evidence for a finite-size behavior of the system arises from the
tail in $\lambda_{ab}^{-2}\left( T\right)$ observed in the
vicinity of $T_{c}$. We have shown that the scaling properties of
the tail are fully consistent with a finite-size effect, arising
from domains with nanoscale size along the $c$-axis. Indeed the
essential characteristics of a finite-size effect, as (i) the
limiting properties of the scaling function $ g(t/|t_{p_{c}}|)$,
(ii) the existence of an inflection point in $\lambda ^{-2}(T)$
and (iii) the extremum in $d\lambda ^{-2}(T)/dT$ at $T_{p_{c}}$
have been verified. Even though YBa$_{2}$Cu$_{4}$O$_{8}$ is {\it
stoichiometric} we have shown that the size of the domains along
the $c$-axis is of nanoscale only, ranging from 33.2~\AA \  to
28.9~\AA \  at pressures from 0.5 to 11.5~kbar. This raises
serious doubts on the existence of macroscopic phase coherent
superconductivity. Contrariwise this does not exclude a
percolative resistive superconductor to normal state transition,
when the  superconducting domains percolate. Indeed, a granular
superconductor is usually characterized by two
parameters:\cite{deutscher} the first is the domain's size and the
second the coupling between them. The coupling is accomplished
randomly with a temperature dependent probability. Such a
mechanism is a percolation process; at the temperature where the
coupling probability is equal to the percolation threshold, an
infinite cluster of coupled superconducting grains is formed. This
suggests that resistive bulk superconductivity is achieved by a
percolative process, while the phase coherent superconducting
properties and the spatial extent of the domains can be probed by
thermal fluctuations and finite-size effects.

The authors are grateful to J.~Roos for stimulating discussions,
K.~Conder for help during sample preparation and T.~Str\"assle for
providing pressure cell for magnetization measurements.
% and M.~Mali for providing sample.
This work was supported by the Swiss
National Science Foundation and by the NCCR program
\textit{Materials with Novel Electronic Properties} (MaNEP)
sponsored by the Swiss National Science Foundation.

\end{document}